\documentclass[11pt]{article}
\usepackage[margin=1in]{geometry}
\usepackage{graphicx}
\usepackage{natbib}
\usepackage{footnote,enumerate,amsmath,amssymb,amsfonts,amsthm,subcaption,hyperref}
\usepackage[linesnumbered,ruled]{algorithm2e}

\newcommand{\R}{{\mathbb R}}
\newcommand{\B}{{\mathbb B}}

\begin{document}
\title{Optimizing the spin reversal transform on the D-Wave 2000Q}
\author{Elijah Pelofske \and Georg Hahn \and Hristo Djidjev}
\date{Los Alamos National Laboratory}
\maketitle

\begin{abstract}
    Commercial quantum annealers from D-Wave Systems make it possible to obtain approximate solutions of high quality for certain NP-hard problems in nearly constant time. Before solving a problem on D-Wave, several pre-processing methods can be applied, one of them being the so-called spin reversal or gauge transform. The spin reversal transform flips the sign of selected variables and coefficients of the Ising or QUBO (quadratic unconstrained binary optimization) representation of the problem that D-Wave minimizes. The spin reversal transform leaves the ground state of the Ising model invariant, but can average out the biases induced through analog and systematic errors on the device, thus improving the quality of the solution that D-Wave returns. This work investigates the effectiveness of the spin reversal transform for D-Wave 2000Q. We consider two important NP-hard problems, the Maximum Clique and the Minimum Vertex Cover problems, and  show on a variety of input problem graphs that using the spin reversal transform can yield substantial improvements in solution quality. In contrast to the native spin reversal built into D-Wave, we consider more general ways to reverse individual spins and we investigate the dependence on the problem type, on the spin reversal probability, and possible advantages of carrying out reversals on the qubit instead of the chain level. Most importantly, for a given individual problem, we use our findings to optimize the spin reversal transform using a genetic optimization algorithm.
\end{abstract}

\section{Introduction}
\label{sec:intro}
Commercial quantum computers from D-Wave Systems Inc.~\cite{D-WaveSystems2000QuantumToday} are designed to approximately solve NP-hard optimization problems that can be expressed as the minimization of a QUBO (quadratic unconstrained binary optimization) or an Ising problem, given by
\begin{align}
    H(x_1,\ldots,x_n) = \sum_{i=1}^n a_i x_i + \sum_{i<j} a_{ij} x_i x_j,
    \label{eq:hamiltonian}
\end{align}
using a process called \textit{quantum annealing}. In \eqref{eq:hamiltonian}, the coefficients $a_i \in \R$ are the linear weights, and $a_{ij} \in \R$ are the quadratic couplers defining the problem for $i,j \in \{1,\ldots,n\}$. If $x_i \in \{0,1\}$, \eqref{eq:hamiltonian} is called a QUBO problem. If $x_i \in \{-1,+1\}$, \eqref{eq:hamiltonian} is called an Ising problem. Both the QUBO and Ising formulations are equivalent \cite{Djidjev2016EfficientAnnealing}. Many important NP-hard problems can be expressed as the minimization of a quadratic function of the form \eqref{eq:hamiltonian}, see \cite{Lucas2014}. We denote the number of variables in the Ising model as $n$ throughout the remainder of the article.

When submitting a problem of the form \eqref{eq:hamiltonian} to the D-Wave annealer, it is preprocessed in two ways. First, the qubits are arranged on the quantum annealer in a particular graph structure, called the \textit{Chimera} graph, consisting of a lattice of bipartite cells on the physical chip \cite{Chapuis2017}. However, the connectivity structure of a QUBO or Ising model, that is its nonzero couplers in \eqref{eq:hamiltonian}, does not necessarily match the connectivity of the Chimera graph. In order to alleviate this issue, a \textit{minor embedding} of the QUBO or Ising connectivity to the Chimera graph can be computed. In such an embedding, several physical qubits are identified to act as one logical qubit in \eqref{eq:hamiltonian}, often severely limiting the number of available qubits. The set of physical qubits on the chip representing a logical qubit (problem variable) is called a \textit{chain}. Second, although $a_i, a_{ij} \in \R$ can be arbitrary, they are rescaled to $a_i \in [-2,2]$ and $a_{ij} \in [-1,1]$ and subsequently converted to analog currents on the chip using an 8-bit digital-to-analog converter.

During the annealing, \textit{leakage} on the physical chip from the coupler $a_{ij}$ ($i,j \in \{1,\ldots,n\}$) can alter the linear weights $a_i$ and $a_j$ \cite{dwave_gauge}. This effect is reported to be often more serious for chained qubits. Moreover, during the process of digital-to-analog conversion, the linear weights $a_i$ will not be perfectly mapped to currents on the chip, and thus biased to the above or below.

The so-called \textit{spin reversal} or \textit{gauge} transform is a simple way to alleviate this issue for Ising problems. The spin reversal is based on the observation that, although theoretically quantum annealing is invariant under a gauge transformation, the calibration of the D-Wave device is not perfect and breaks the gauge symmetry. This implies that, indeed, spin reversed Ising systems realize slightly different systems on the annealer, yielding different results which can be averaged. To apply the spin reversal, we flip the sign of an arbitrary number of variables and coefficients of the Ising problem, thus resulting in a reinterpretation of an \textit{up} as a \textit{down} spin and vice versa. This leaves the ground state of \eqref{eq:hamiltonian} invariant, but has the potential to reduce analog and systematic errors on the device as described earlier (by averaging them out), thus improving the quality of the solution.

In particular, to transform the qubit $x_i$ from $-1$ to $+1$, we define a new function $H'$ with $a_i' \rightarrow -a_i$ as well as $a_{ij}' \rightarrow -a_{ij}$ and $a_{ji}' \rightarrow -a_{ji}$ for all $j \in \{1,\ldots,n\}$. We observe that the ground state energies of $H$ and $H'$ are identical, and that the minimum of $H'$ is the minimum of $H$ with the $i$-th variable having a flipped sign. As reported in \cite{dwave_gauge}, reversing too few spins leaves the Ising model almost unchanged, whereas applying the spin reversal transform to too many qubits likely results in many pairs of connected qubits being transformed, thus effectively leaving the corresponding quadratic couplers unchanged. In both cases, the spin reversal transform might only have little effect.

It is important to note that the spin reversal transform can be applied on two different levels: After embedding the Ising model to be solved onto the D-Wave architecture, the actual embedded problem that D-Wave solves is read and the spin reversal is applied to any qubit independently -- this is referred to in the remainder of the article as \textit{spin reversal on the qubit level}. Second, we can apply the spin reversal in such a way that the physical qubits in a chain (representing one logical qubit) are all either spin reversed or all left unchanged -- we refer to this technique in the remainder of the article as \textit{spin reversal on the chain level}.

The SAPI interface of D-Wave allows us to apply a built-in (simple) form of the spin reversal transform. It is controlled through the parameter \textit{num\_spin\_reversal\_transforms} ($=N_s$) in the function \textit{solve\_ising}. In connection with the parameter \textit{num\_reads} ($=N_r$) which specifies the total number of anneal readouts, D-Wave will generate $N_s$ spin reversed Ising problems and obtain $N_r/N_s$ readouts for each. The spin reserved Ising models are obtained by flipping each qubit independently with probability roughly $0.5$.

In this work, we aim to assess the effectiveness of the spin reversal transform in a more general way. The main contribution of this article is twofold. First, we evaluate the performance of the spin reversal transform: In particular, we apply it to both the raw Ising formulation and the embedded problem. We are furthermore interested in its performance as a function of the probability of flipping a single qubit. Lastly, we assess its performance on two different NP-hard graph problems, the Maximum Clique problem and the Minimum Vertex Cover problem, see \cite{Lucas2014}, which will be introduced later.

Second, we aim to use our findings to optimize the spin reversal transform in practice. For this we employ a genetic optimization algorithm which, for a given problem instance given as an Ising model, finds the set of qubits on which the spin reversal transform is most effective. For this we consider a separate binary indicator for each qubit (reversed or not reversed), and optimize over all $n$ indicators to find the best configuration.

In this article, we  provide a rigorous assessment of the effectiveness of the spin reversal transform, which to the best of our knowledge has not been presented in the literature previously. Existing work published in the literature does employ spin reversals, yet only as a tool to possibly enhance solutions, and only using the in-built D-Wave implementation. In \cite{ttt} the authors define a new metric, the \textit{time-to-target} metric, as the time needed by classical solvers to match the results of a quantum annealer: for their experiments, the authors employ D-Wave's native gauge transform, but it is left unclear what the contribution of the transform to the solution quality is. In \cite{Prudenz}, the author primarily evaluates several techniques to allocate weights to chains of qubits, as well as two ways of determining the final value of a chain. The effect of four spin reversals is also considered briefly, however no further statement is made on the spin reversal transform apart from the fact that it shows a slight performance gain on certain systems. In \cite{KingMcGeoch}, the authors conclude that gauge transformations are more effective on difficult problems. However, in \cite{Boixo2014}, the authors demonstrate that significant correlations exist between different gauge transforms.

The article is structured as follows. Section~\ref{sec:gauge} describes the spin reversal transform in detail, and presents a genetic algorithm to attempt to solve the optimization over all possible spin reversal transformations. Section~\ref{sec:experiments} presents simulation results for the Erd\H{o}s-R\'enyi graph family, both as a function of the input graph density as well as of the probability of the spin reversal transform, and for the two aforementioned NP-hard graph problems. We also present results highlighting the dependence of the genetic optimization algorithm on its parameters, and show how the effectiveness of the spin reversal transform can be considerably increased in comparison to the D-Wave transformation over the course of only a few 'genetic' generations. The article concludes with a discussion in Section~\ref{sec:discussion}.

\section{The Gauge Transform}
\label{sec:gauge}
This section describes the setup we employ to apply the spin reversal transform.

\subsection{Spin reversal on the qubit level}
\label{sec:spin_qubit}
Given an input Ising model of type \eqref{eq:hamiltonian}, we embed it onto the D-Wave Chimera graph first.

Before starting the annealing process, however, we read the embedded Ising model from the D-Wave connectivity graph. The embedded problem typically consists of more variables, precisely the physical chain qubits representing the logical qubits.

Given the embedded Ising model on the D-Wave chip with $n$ qubits, we can reverse spins in the following way. Select a set $I \subseteq \{1,\ldots,n\}$ of qubit indices to be switched. In our experiments, we will generate the set $I$ by adding each index in $\{1,\ldots,n\}$ to it independently with a given probability $p_s$.

We then sequentially select one $i \in I$ at a time, and set $a_i' := -a_i$. Furthermore, we set $a_{ij}' := -a_{ij}$ and $a_{ji} := -a_{ji}$ for all $j \in \{1,\ldots,n\}$. After having applied this procedure for all $i \in I$, we use the new sets of linear weights $\{a_i'\}$ and quadratic couplers $\{a_{ij}'\}$ to form a new Ising problem $H'$.

The new $H'$ is then embedded onto the D-Wave chip (instead of the originally embedded problem) and solved.

\subsection{Spin reversal on the chain level}
\label{sec:spin_chain}
Another way of applying the spin reversal is at the chain level. In contrast to Section~\ref{sec:spin_qubit}, we read the embedded problem and  switch the signs of either all physical qubits in a chain that encode one logical qubit, or of none of them.

\subsection{Optimizing the spin reversal transform}
\label{sec:genetic}

\begin{algorithm}[t]
    \caption{Genetic algorithm for spin reversal tuning\label{algo:genetic}}
    \SetKwInOut{Input}{input}
    \SetKwInOut{Output}{output}
    \SetKwFor{Repeat}{repeat}{}{end}
    \Input{$H$, $N$, $p_\text{spin}$, $p_\text{mat}$, $p_\text{mut}$, $R$, $N_a$\;}
    \Output{final population of spin reversal vectors\;}
    $n \leftarrow$ number of variables in $H$\;
    $S \leftarrow \{s_1,\ldots,s_N: s_i \in \B_n^{p_\text{spin}} \}$\;
    \label{line:init}
    \For{$r \leftarrow 1$ \KwTo $R$}{
        \For{$s \in S$}{
            \label{line:for}
            $H' \leftarrow$ reverse spins $s$ in $H$\;
            Request $N_a$ anneals for $H'$ on D-Wave and store minimal energy among those in $e_s$\;
        }
        $E \leftarrow \{e_s:s \in S\}$\;
        $E_0 \leftarrow$ proportion $p_\text{mat}$ of lowest energies in $E$\;
        $S_0 \leftarrow \{s \in S: e_s \in E_0\}$\;
        $w_0 \leftarrow$ $\arg\max\{e_s:s \in S\}$\;
        \If{$r>1$}{
            Replace $w_0$ in $S$ with $w_1$\;
        }
        $S_1 \leftarrow \emptyset$\;
        \Repeat{$N$ \textnormal{\textbf{times}}}{
            \label{line:crossover}
            Draw two random $s_1,s_2 \in S_0$ and combine bits randomly with probability $0.5$; store result in $S_1$\;
        }
        $S_2 \leftarrow \emptyset$\;
        \For{$s \in S_1$}{
            \label{line:mutation}
            Flip each bit in $s$ independently with probability $p_\text{mut}$ and store the mutated $s$ in $S_2$\;
        }
        $S \leftarrow S_2$\;
        $w_1 \leftarrow$ $\arg\min\{e_s:s \in S\}$\;\label{line:bestindividual}
    }
    \Return{$S$}\;
\end{algorithm}

Algorithm~\ref{algo:genetic} presents the genetic algorithm we use to optimize the spin reversal transform for the D-Wave annealer. Let an input Ising model $H$ with $n$ variables be given which we aim to solve on D-Wave. In the following, we characterize a spin reversal through a boolean vector of length $n$, where each \textit{True} entry encodes that the corresponding variable is reversed.

Using Algorithm~\ref{algo:genetic}, we aim to find the spin reversal transform (the boolean vector) among all $2^n$ possible binary vectors that yields the minimal average energy on D-Wave (within a pre-specified number of $N_a$ anneals).

We use a genetic algorithm to optimize over generations of boolean vectors (or bitstrings) of length $n$ which are interpreted as spin reversals in the aforementioned sense. Algorithm~\ref{algo:genetic} works as follows: First, in line~\ref{line:init}, we draw $N$ random bitstrings from $\B_n^{p_\text{spin}}$ having length $n$ and probability $p_\text{spin}$ for an entry \textit{True}. Those are stored in a set $S$.

Next, the initial population in $S$ is evaluated with regards to their quality of solution. For this, we apply the spin reversal to the variables indicated in each $s \in S$, resulting in a new Ising model $H'$, perform $N_a$ anneals for $H'$ on D-Wave, and record the minimal energy $e_s$ obtained in this way for $s$. All those energies are stored in a set $E$. Afterwards, we compute a subset $E_0$ of $E$ corresponding to the proportion $p_\text{mat}$ of lowest (i.e., best) energies. We store the boolean strings that correspond to the energies in $E_0$ in a set $S_0$. Additionally, we record the bitstring $w_0$ leading to the highest (i.e., worst) energy.

Next, for all generations after the first one, the \textit{worst} individual $w_0$ is replaced by the individual $w_1$ that resulted in the minimal (i.e., best) energy in the previous generation (see line~\ref{line:bestindividual}), thereby ensuring that the current global minimum solution found is never lost.

Next, we carry out a \textit{crossover} operation on the population in line~\ref{line:crossover}. For this, we take two arbitrary $s_1,s_2 \in S_0$ and combine them into a new boolean vector by choosing its entries independently from $s_1$ and $s_2$ with probability $0.5$. The boolean vector obtained in this way is added to a set $S_1$ which was initialized as $S_1=\emptyset$. This is repeated $N$ times in order to leave the population size invariant in each generation.

Finally, a \textit{mutation} step is applied in line~\ref{line:mutation}. We flip the entries of each $s \in S_1$ independently with a low probability $p_\text{mut}$ and store the resulting $s$ in a set $S_2$ (likewise initialized with $S_2=\emptyset$). After mutations have been applied to all $s \in S_1$, we restart the evaluation of the population with the updated set $S_2$. In preparation for the next iteration, in line~\ref{line:bestindividual}, the bitstring $w_1$ is recorded with lowest (i.e., best) energy in population $S$.

We create $R$ generations in this way.

After termination of Algorithm~\ref{algo:genetic}, we are left with the last population of boolean vectors which encode different spin reversals. In order to choose our best candidate for applying the spin reversal to the Ising model $H$, we determine their minimal energies one last time: We apply each to $H$ (resulting in a new $H'$ as before), record the lowest minimal energy for $H'$ in $N_a$ anneals on D-Wave, and return the boolean vector among the entire population yielding the minimum energy solution.

\section{Experiments}
\label{sec:experiments}
In all the experiments that follow, we consistently employed D-Wave's default parameters with the exception of annealing time, which we set to 1 microsecond.

\subsection{Evaluation of the spin reversal transform}
\label{sec:results_gauge}
In the following two subsections, we evaluate the spin reversal transform on both the qubit and the chain level.

The test graphs we employ are $G(|V|,p_G)$ Erd\H{o}s--R\'enyi graphs with $|V|=64$ vertices and edge probability $p_G \in \{0.1, 0.3, 0.5, 0.7, 0.9\}$. The reason for using $|V|=64$ vertices stems from the fact that this is the largest size of an arbitrary graph that can be embedded onto  D-Wave 2000Q, see \cite{Pelofske2019,Pelofske2019vertex}.

In order to generate an Ising model to solve for each test graph, we consider two NP-hard graph problems: the Maximum Clique problem, which asks for a maximum fully-connected set of vertices of a graph, and the Minimum Vertex Cover problem, which asks for a minimum set of vertices in a graph such that each edge is incident to at least one vertex from that set. The QUBO/Ising formulations of those two problems in the form of \eqref{eq:hamiltonian} can be found in \cite{Lucas2014}.

The spin reversal transform requires the index set $I$ which specifies the qubits to be spin-reversed, see Section~\ref{sec:spin_qubit}. In the simulations of this section we generate the index sets with varying probabilities $p_s \in \{ 0.01, 0.05, 0.1, 0.2, \ldots, 0.8, 0.9, 0.95, 0.99 \}$, that is for each element in $\{1,\ldots,n\}$ we decide with an independent Bernoulli trial of probability $p_s$ whether it should be included in $I$ or not (where $n$ is the number of variables in the Ising model, see \eqref{eq:hamiltonian}).

For each value of $p_s$, we generate $50$ graphs, apply the two Ising model formulations for the Maximum Clique and Minimum Vertex Cover problems, and solve them using the following two techniques using $N_a=1000$ anneals on D-Wave: (a) we solve with D-Wave's spin reversal parameter using $N_s=10$ (see Section~\ref{sec:intro}) -- we will refer to this technique as \textit{D-Wave's native spin reversal}; (b) with spin reversal on the qubit level; (c) with spin reversal on the chain level. For every batch of anneals, we report the best solution found as the mean among the $1\%$ lowest energy solutions on D-Wave. This is to report a more robust measurement than merely recording the best solution found.

In this subsection, all lines in the figures are averages over $50$ repetitions. For each repetition, we regenerate both the input graph, the Ising model, and its embedding on D-Wave.

\subsubsection{Results for spin reversal on the qubit level}
\label{sec:results_qubit}
\begin{figure}
    \centering
    \includegraphics[width=0.49\textwidth]{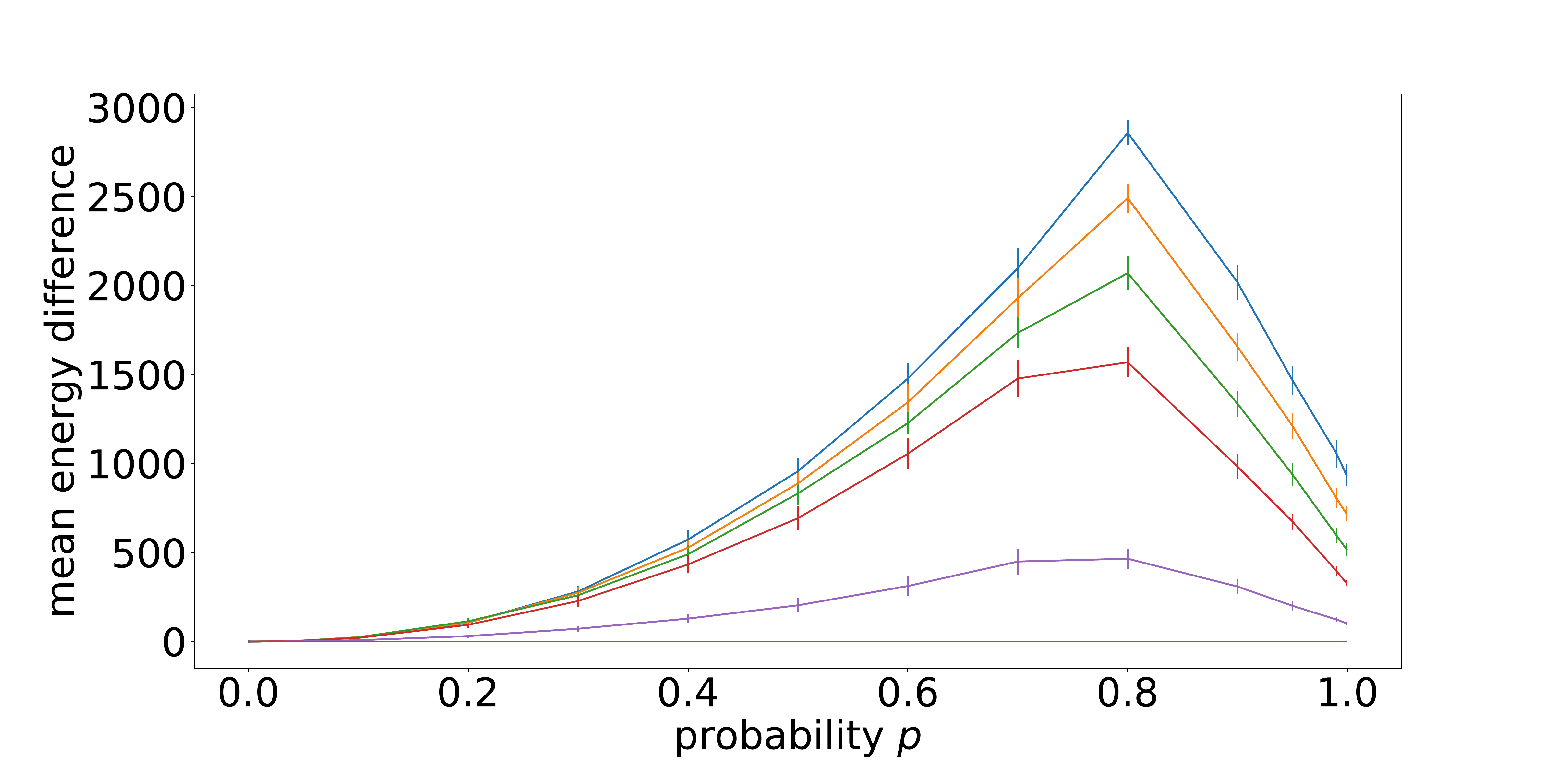}~
    \includegraphics[width=0.49\textwidth]{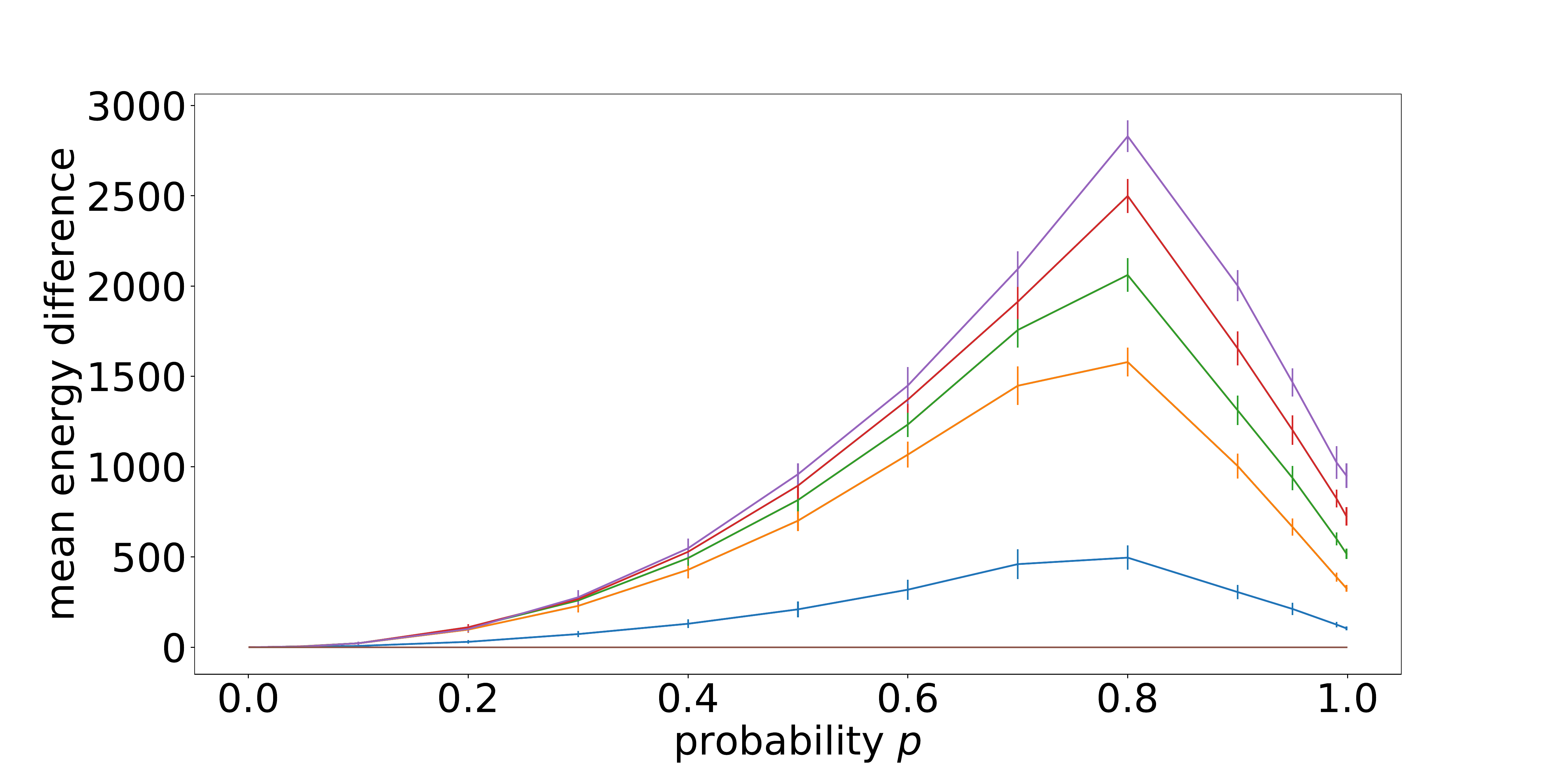}
    \caption{Mean energy difference of the spin reversal on qubit level of Section~\ref{sec:spin_qubit} to the minimum energy found from a normal annealing schedule as a function of $p_s$. Test graphs generated with edge probabilities $0.1$ (blue), $0.3$ (yellow), $0.5$ (green), $0.7$ (red) and $0.9$ (magenta), brown line is the D-Wave native spin reversal. Maximum Clique (left) and Minimum Vertex Cover (right) problems.\label{fig:spin_reversal_qubit}}
\end{figure}
Figure~\ref{fig:spin_reversal_qubit} shows experimental results for a comparison of D-Wave's standard anneal to a general spin reversal approach of Section~\ref{sec:spin_qubit} as a function of the spin reversal probability $p_s$ ($x$-axis). We use test graphs of varying density (blue to magenta) applied to the Maximum Clique (left) and Minimum Vertex Cover (right) problems. The two plots display the mean energy difference to D-Wave's standard anneal on the $y$-axis. We also observe that the mean energy difference between D-Wave standard annealing and D-Wave's built in spin reversal is quite small. We display error bars of one standard deviation for each measurement.

The figure shows that for both the Maximum Clique and the Minimum Vertex Cover problem, the approach of Section~\ref{sec:spin_qubit} always yields a solution of higher energy (worse quality) than the one using D-Wave's normal annealing. 

The curves in Figure~\ref{fig:spin_reversal_qubit} exhibit a maximum (which happens to occur around $p=0.8$): Such a maximum is to be expected, since both very low and very high probabilities leave an Ising problem essentially unchanged (see Section~\ref{sec:intro}).

\subsubsection{Results for spin reversal on the chain level}
\label{sec:results_chain}
\begin{figure}
    \centering
    \includegraphics[width=0.49\textwidth]{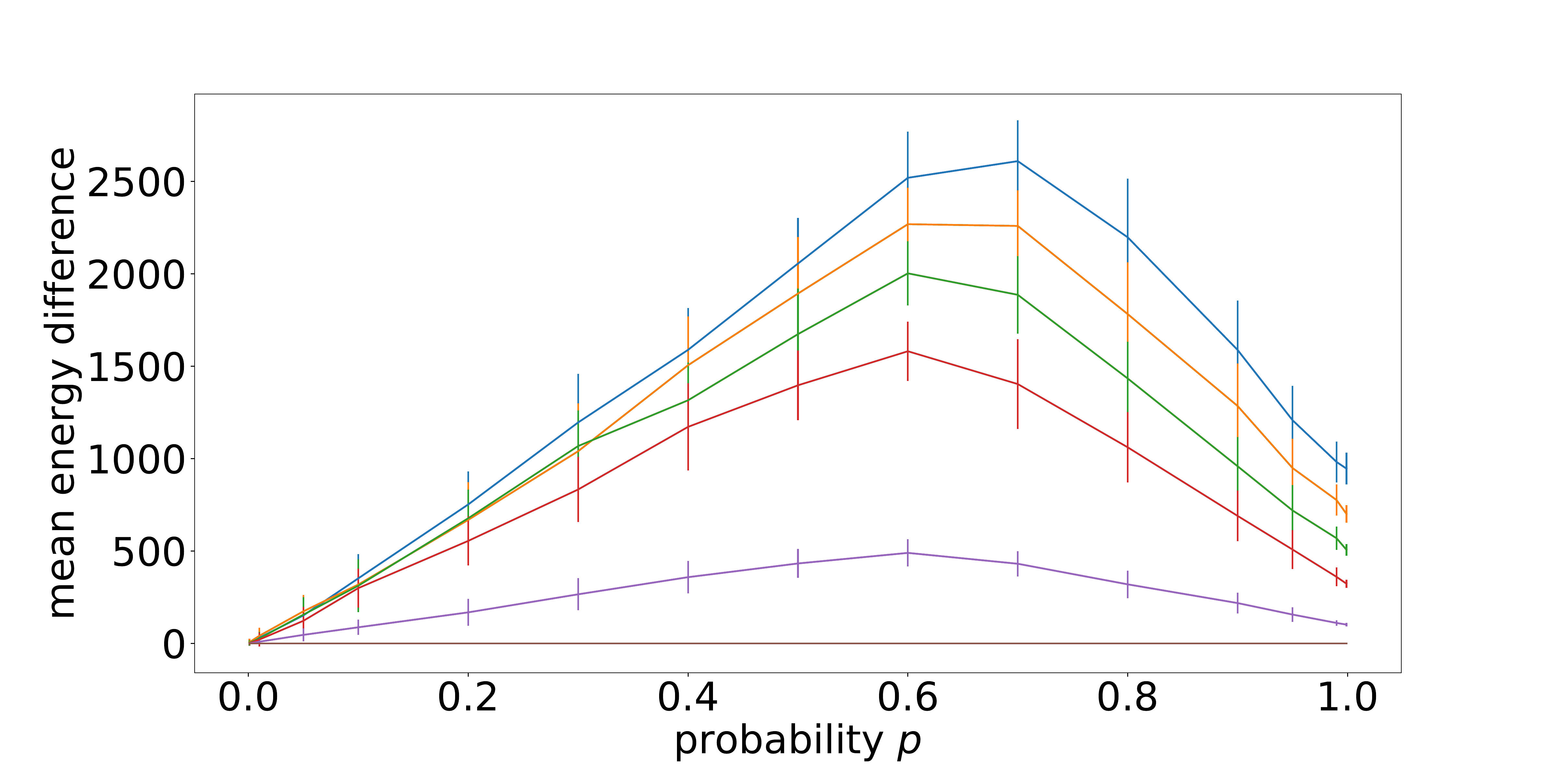}~
    \includegraphics[width=0.49\textwidth]{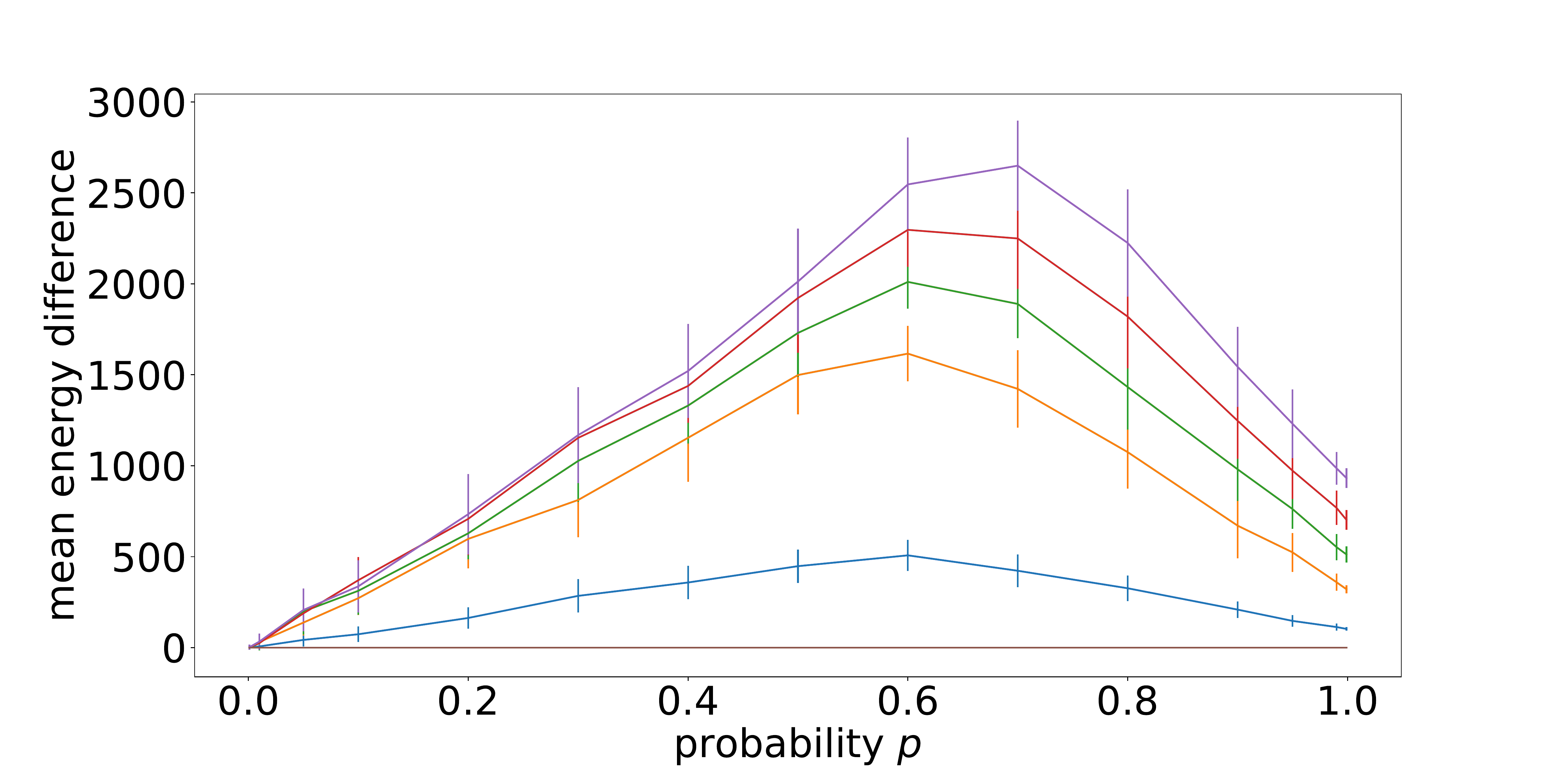}
    \caption{Mean energy difference of the minimum energy found from spin reversal on chain level of Section~\ref{sec:spin_chain} to the minimum energy obtained with D-Wave's standard anneal as a function of $p_s$. Test graphs generated with edge probabilities $0.1$ (blue), $0.3$ (yellow), $0.5$ (green), $0.7$ (red) and $0.9$ (magenta), brown line is the D-Wave native spin reversal. Maximum Clique (left) and Minimum Vertex Cover (right) problems.\label{fig:spin_reversal_chain}}
\end{figure}
Figure~\ref{fig:spin_reversal_chain} shows results of a comparison of D-Wave's standard anneal to the spin reversal on the chain level described in Section~\ref{sec:spin_chain}. As in Section~\ref{sec:results_qubit}, our approach of Section~\ref{sec:spin_chain} is performing worse than D-Wave's  standard annealing in all cases.

\subsection{Results for our genetic algorithm}
\label{sec:results_genetic}
\begin{figure}
    \centering
    \includegraphics[width=0.49\textwidth]{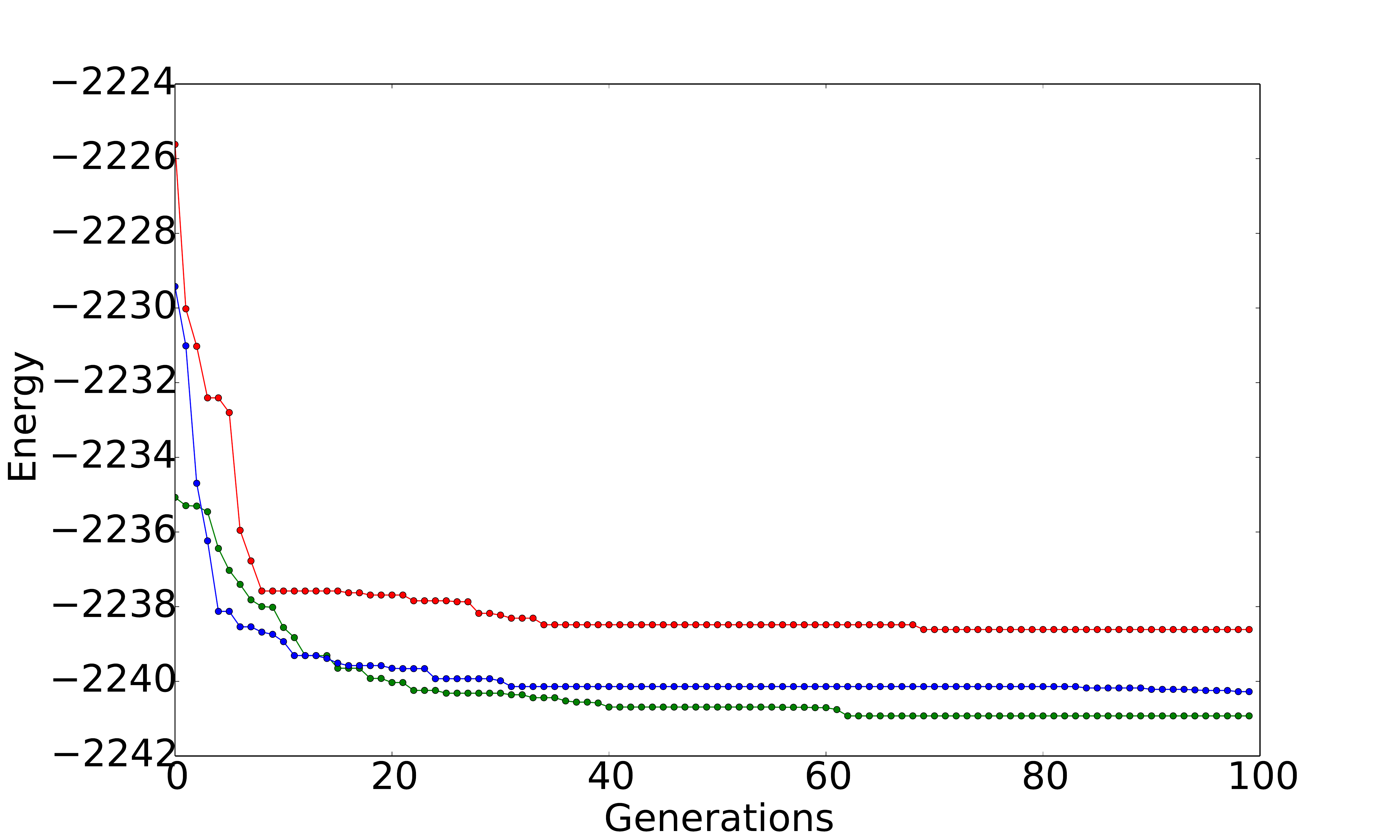}~
    \includegraphics[width=0.49\textwidth]{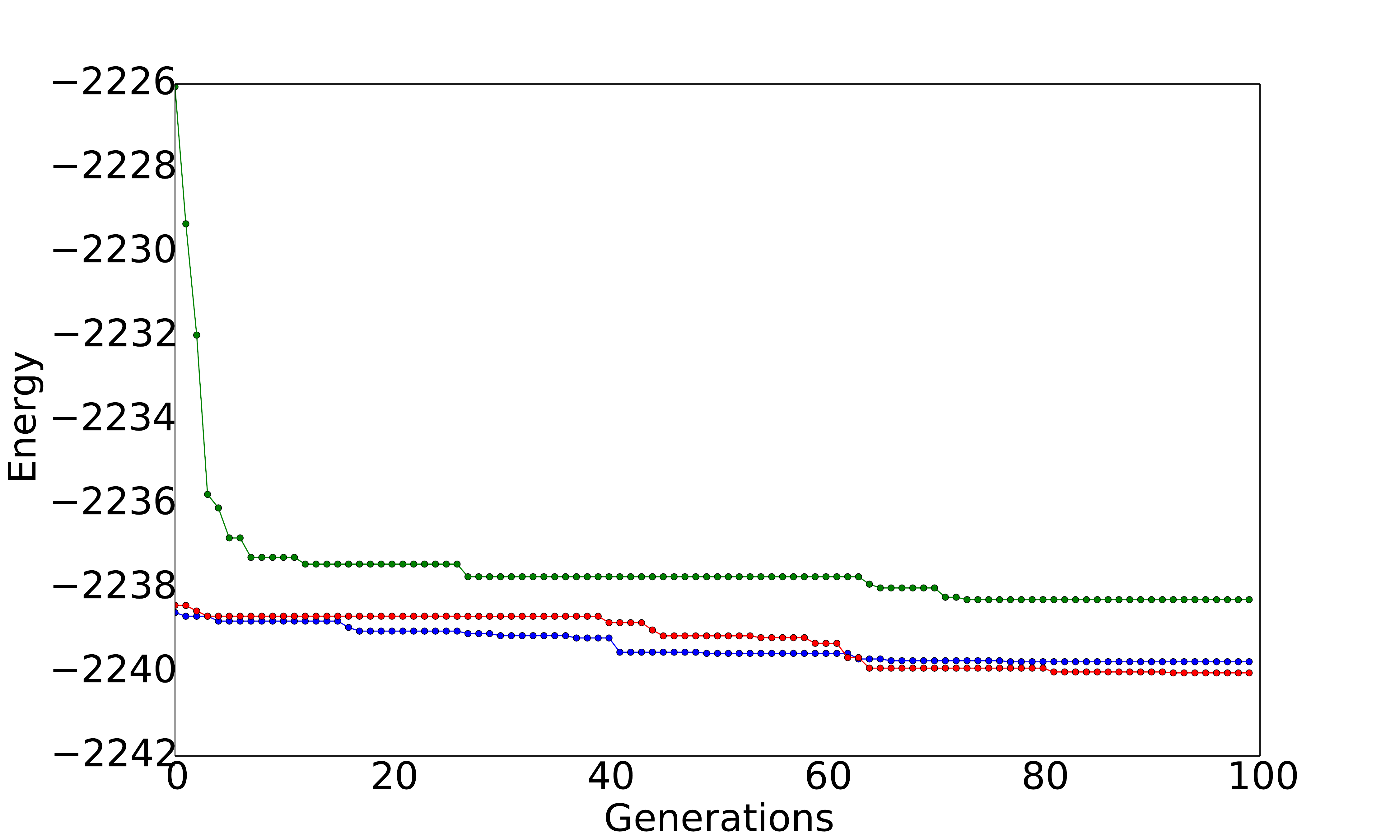}\\
    \includegraphics[width=0.49\textwidth]{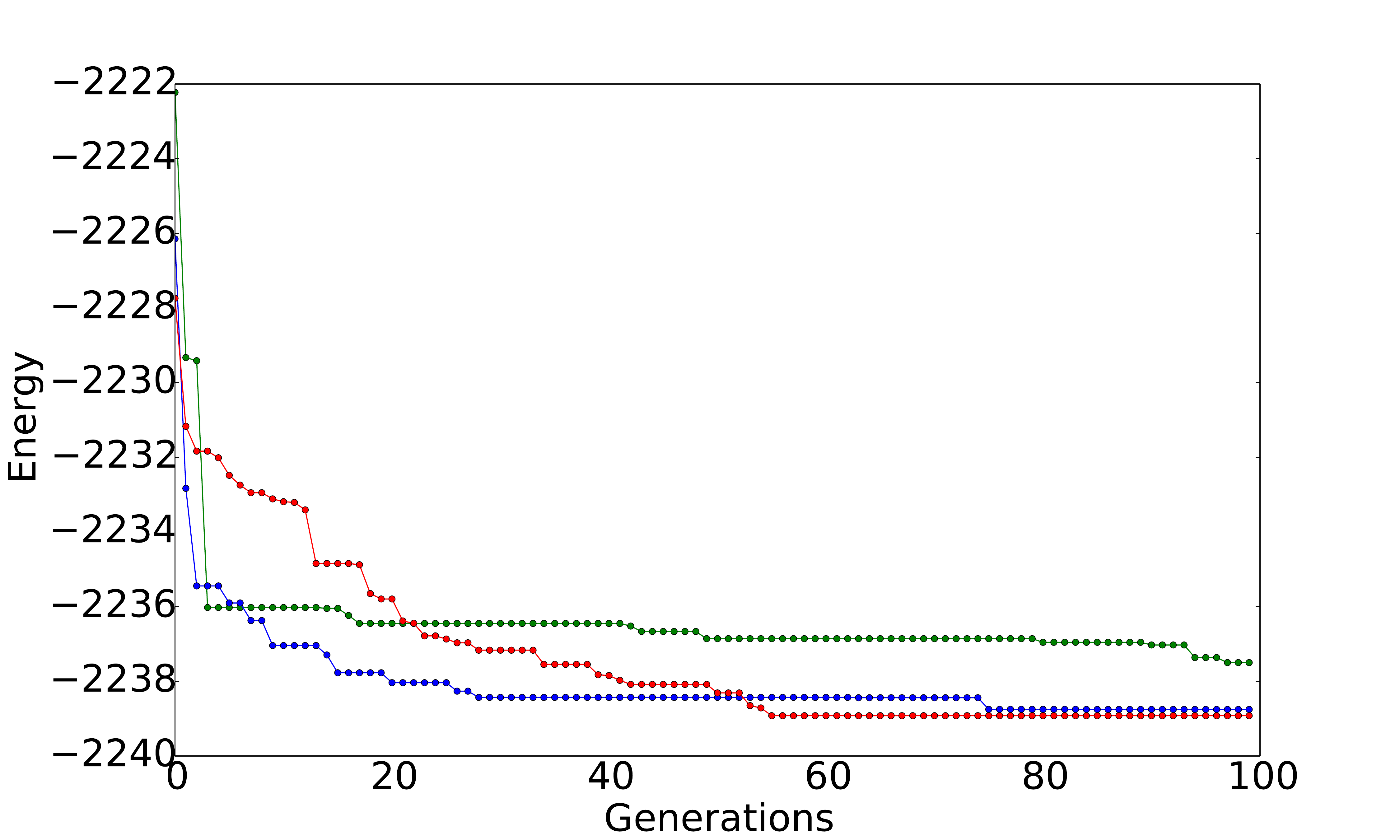}~
    \includegraphics[width=0.49\textwidth]{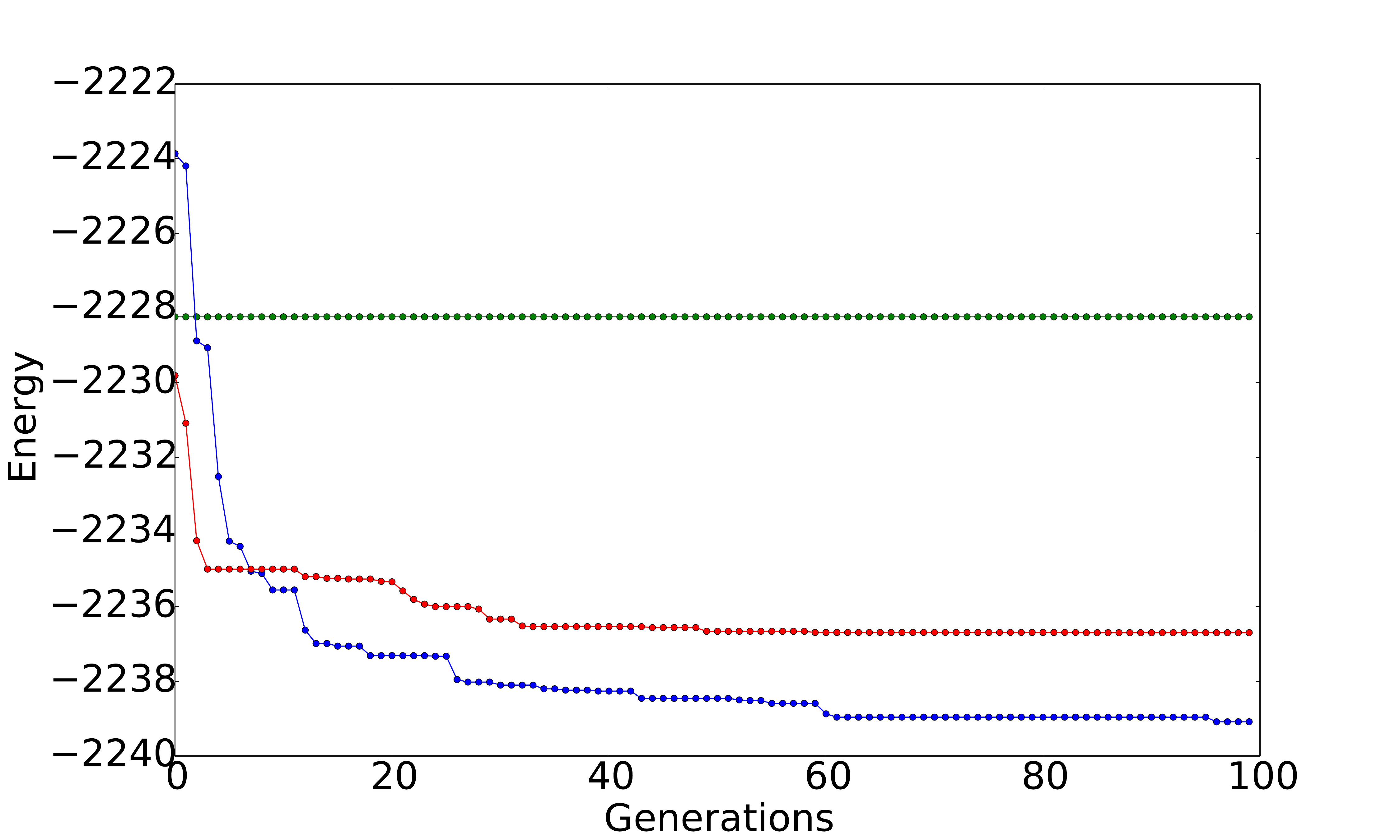}
    \caption{Dependence of Algorithm~\ref{algo:genetic} on its parameters $N \in \{20,50,80\}$ (top left), $p_\text{spin} \in \{0.001,0.01,0.1\}$ (top right), $p_\text{mat} \in \{0.1,0.3,0.5\}$ (bottom left) and $p_\text{mut} \in \{0.001,0.01,0.1\}$ (bottom right) when applied to spin reversal at qubit level. We consistently use red for the smallest value, blue for the medium value, and green for the largest parameter value in all subplots.\label{fig:GA_param_chain}}
\end{figure}
This section starts by assessing the dependence of the performance of Algorithm~\ref{algo:genetic} on its parameters. For this, we first define a base set of parameters: $N = 20$, $p_\text{spin} = 0.1$, $p_\text{mat} = 0.1$, and $p_\text{mut} = 0.01$. When varying each of those four parameters, we keep the others fixed at their base values.

Figure~\ref{fig:GA_param_chain} shows results for graphs of density $0.5$ applied to the Ising model of the Maximum Clique problem. The generated graph and the embedding of the Maximum Clique Ising model onto the D-Wave architecture were kept fixed. Algorithm~\ref{algo:genetic} was run over $R=100$ generations. The spin reversal was applied at the qubit level. We assess the dependence on $N \in \{20,50,80\}$ (top left), $p_\text{spin} \in \{0.001,0.01,0.1\}$ (top right), $p_\text{mat} \in \{0.1,0.3,0.5\}$ (bottom left) and $p_\text{mut} \in \{0.001,0.01,0.1\}$ (bottom right). In all subplots of Figure~\ref{fig:GA_param_chain}, the green lines consistently indicate the smallest value of the three investigated choices for each parameter, red indicates the medium value and blue the largest one.

Figure~\ref{fig:GA_param_chain} shows that, as expected, larger values of $N$ result in more diverse populations, which increases the probability of creating population members (in our case, the population members are the bitstrings indicating which qubits need to be spin reversed) resulting in spin transforms yielding a low energy. Likewise, the spin reversal probability $p_\text{spin}$ (used to create the initial population) should be rather low, although high values of $p_\text{spin}$ eventually result in the same low energy solution given enough generations are created. The mutation probability $p_\text{mut}$ should likewise be chosen small, since large values have a tendency to add too much noise to good solutions. The rate $p_\text{mat}$ specifying the proportion of population members with lowest energies after annealing that are used in the crossover stage should similarly be small in order to only pass on the best candidates to the next generation, however if the crossover proportion is too small, diversity among the population becomes too low.

Note that in Figure~\ref{fig:GA_param_chain} (top right) assessing the dependence on the spin reversal probability $p_\text{spin}$, in contrast to the other three plots the blue, red and green lines do not start in the same point in generation zero. This can be explained as follows: Only the parameters $p_\text{spin}$ and $N$ appear in the initialization in line~\ref{line:init} of Algorithm~\ref{algo:genetic}. Whereas for large enough $N$, the lowest energy solution of the initial population stays roughly the same, the spin reversal probability $p_\text{spin}$ affects the population and its overall minimal energies as a whole in generation zero.

We observe that in all subfigures of Figure~\ref{fig:GA_param_chain}, our genetic algorithm converges to nearly the same minimum energy for the best parameter choice, which is to be expected for a good optimization scheme.

Based on Figure~\ref{fig:GA_param_chain}, we will use Algorithm~\ref{algo:genetic} in the remainder of the article with parameters
\begin{align}
    N=80,~p_\text{spin}=0.1,~p_\text{mat}=0.1,~p_\text{mut}=0.01.
    \label{eq:standard_param}
\end{align}

When using Algorithm~\ref{algo:genetic} in connection with spin reversal at the chain level, see Figure~\ref{fig:GA_param_qubit} in Appendix~\ref{sec:genetic_qubit}.

\begin{figure}
    \centering
    \includegraphics[width=0.7\textwidth]{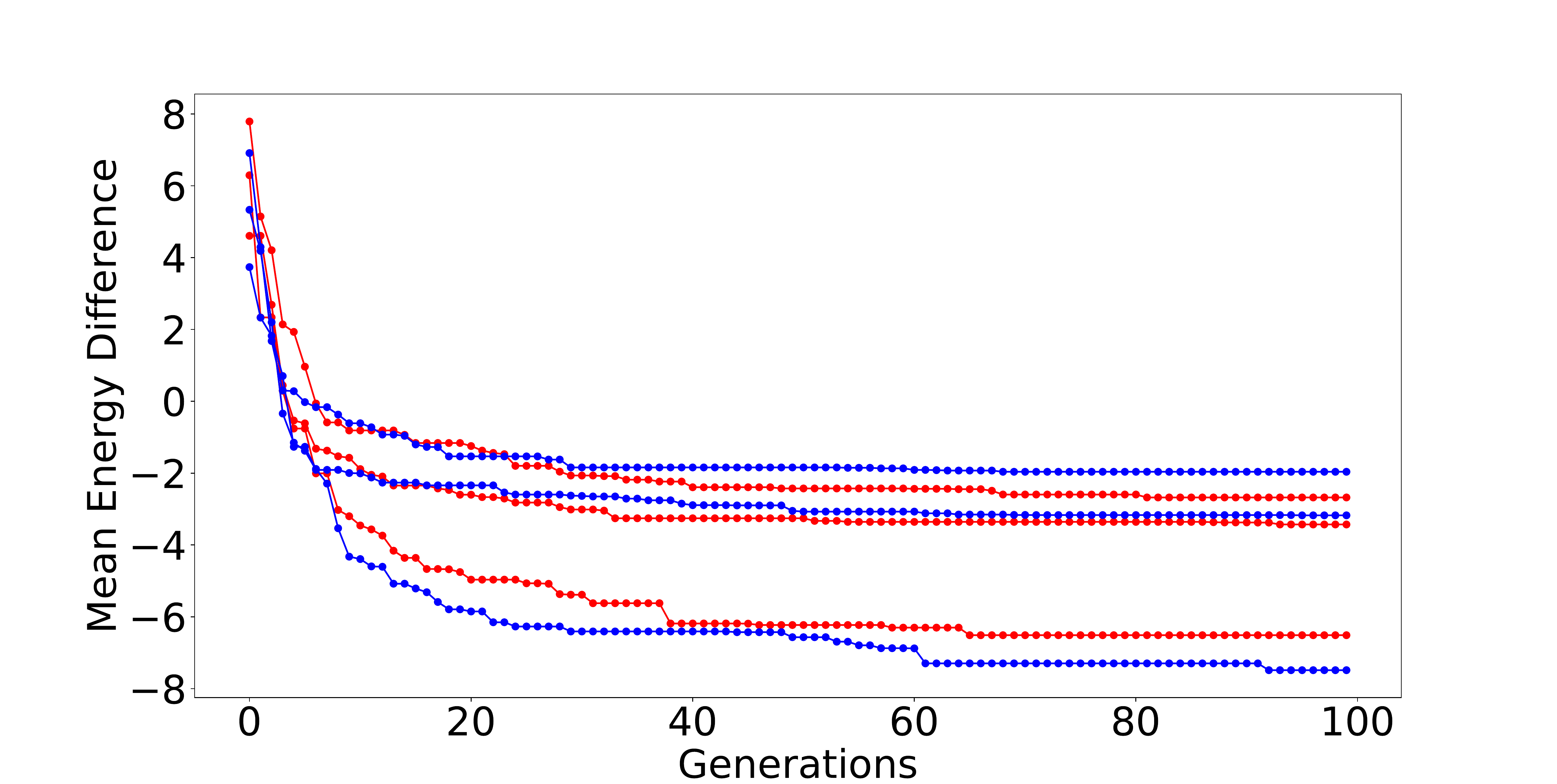}
    \caption{Comparison of the energy difference between D-Wave's native spin reversal ($N_a$=10000, $N_s$=100) and Algorithm~\ref{algo:genetic} as a function of the generation number. Maximum Clique (red) and Minimum Vertex Cover (blue) problems.\label{fig:ga_comp}}
\end{figure}

\begin{figure}
    \centering
    \includegraphics[width=0.7\textwidth]{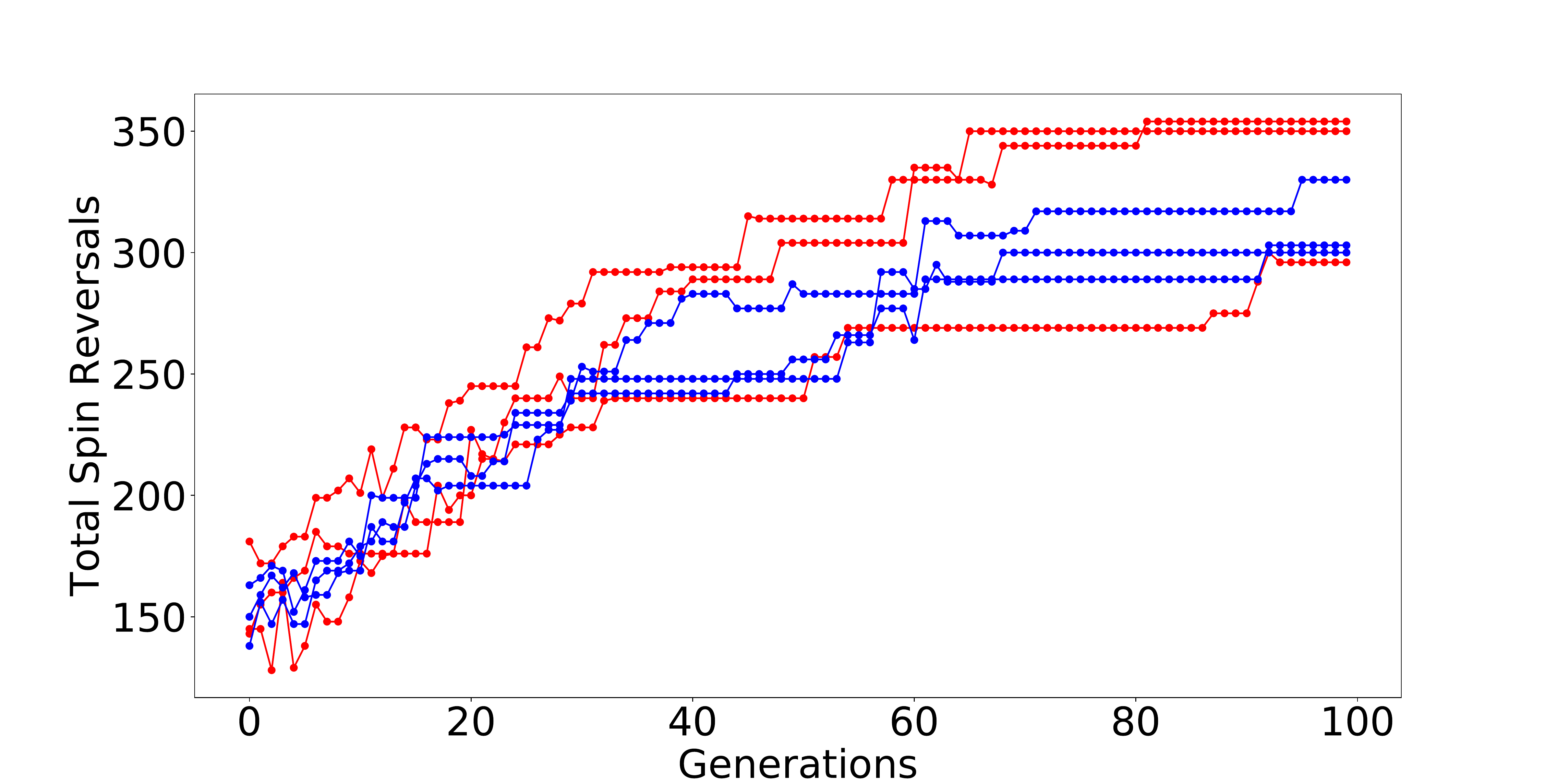}
    \caption{Total number of observed spin reversals in Algorithm~\ref{algo:genetic} as a function of the number of generations. Maximum Clique (red) and Minimum Vertex Cover (blue) problems.\label{fig:ga_number_reversals}}
\end{figure}

Using the parameters given in \eqref{eq:standard_param}, we are able to employ Algorithm~\ref{algo:genetic} to determine a (reasonably) good set of variables to be spin-reversed (with the aim of creating an equivalent Ising model which yields lower minimal energies during annealing).

Figure~\ref{fig:ga_comp} shows simulation results for D-Wave's native spin reversal and six different realizations of Algorithm~\ref{algo:genetic} over $100$ generations. The blue datapoints symbolize results for the Minimum Vertex Cover problem, red stands for the Maximum Clique problem. For each datapoint of the genetic algorithm, we plot the difference of the lowest $1\%$ mean energy (obtained when applying the current best set of variables to be spin reversed to the Ising model of the Maximum Clique problem) to the solution obtained with D-Wave's spin reversal.

We observe that for the set of parameters given in \eqref{eq:standard_param}, after only a handful of generations (around $R>5$), Algorithm~\ref{algo:genetic} has found a combination of variables in the Ising model, which, after spin reversal has been applied to them, yield a modified (but equivalent) Ising model with (substantially) lower energies after annealing than D-Wave's spin reversal.

Figure~\ref{fig:ga_number_reversals} visualizes how the number of spin reversals in the best solution found by Algorithm~\ref{algo:genetic} changes over the course of $R=100$ generations. As before, blue datapoints stand for the Minimum Vertex Cover problem, and red datapoints for the Maximum Clique problem. We observe that the numbers seem to increase steadily, though the increase levels off and, as expected, the number of spin reversals in the best solution starts to plateau towards higher generation numbers.

\subsection{Spatial correlation of spin reversed qubits on the D-Wave chip}
\label{sec:spatial}
\begin{figure}
    \centering\hfill
    \includegraphics[width=0.49\textwidth]{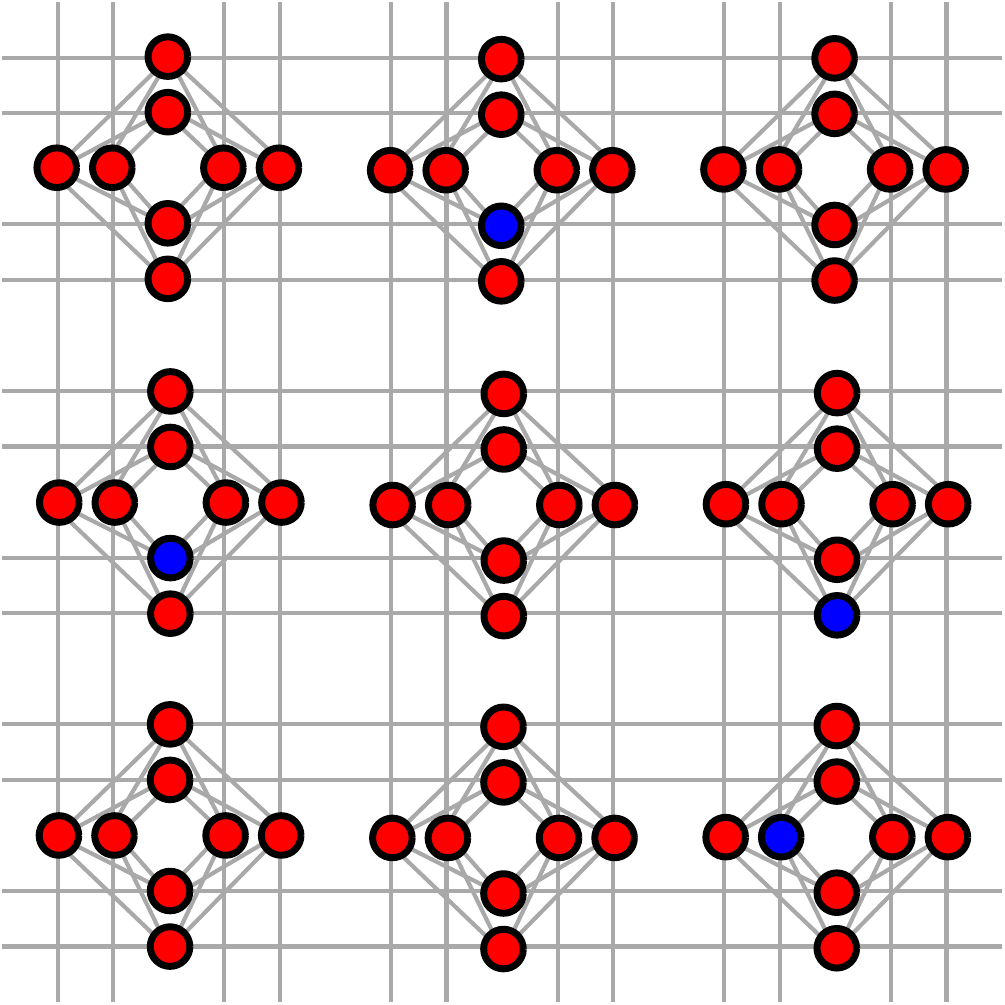}
    \hfill
    \includegraphics[width=0.49\textwidth]{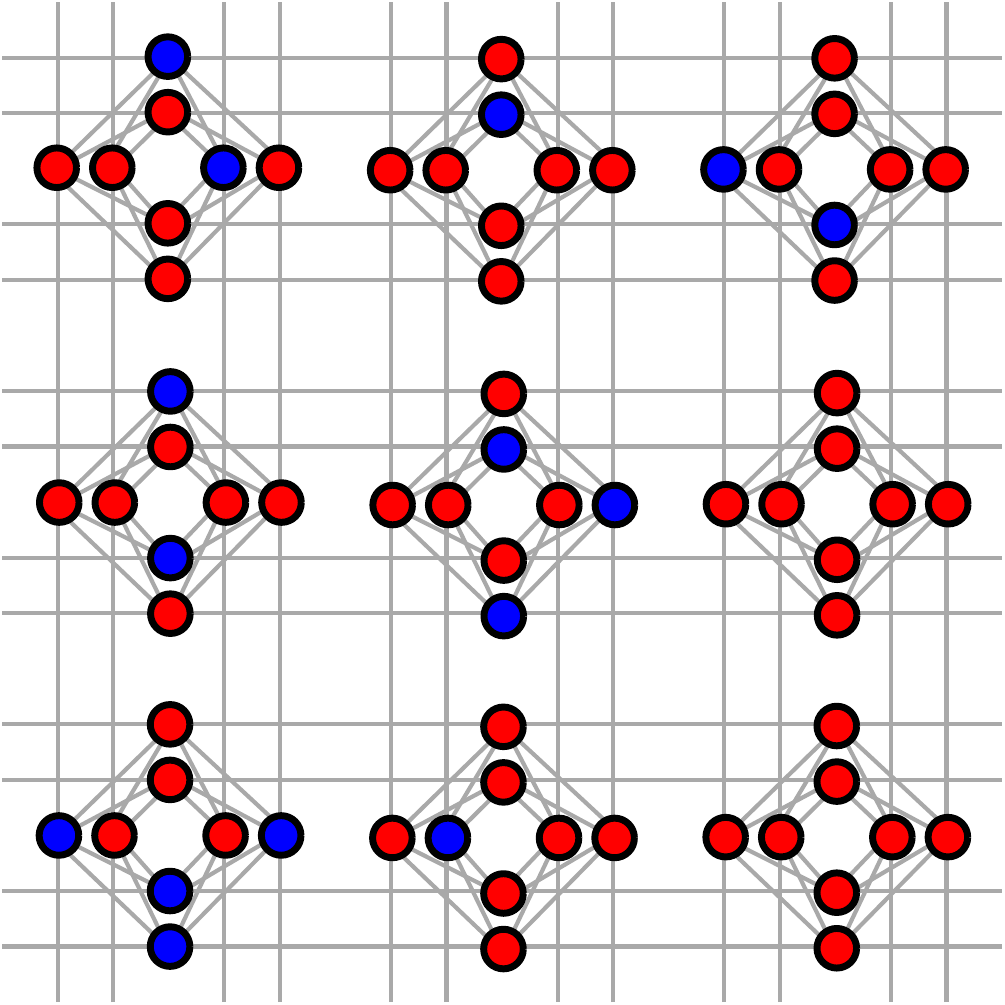}
    \hfill~
    \caption{Section of the qubit connectivity graph of D-Wave~2000Q. Spin reversed qubits displayed in blue and non-reversed qubits in red for the single current best solution of Algorithm~\ref{algo:genetic} at initialization (left) and after $100$ generations (right).\label{fig:spatial}}
\end{figure}
It is of interest to investigate if there is a spatial correlation on the physical D-Wave chip between those qubits for which Algorithm~\ref{algo:genetic} determines that a spin reversal improves the solution quality.

For the Maximum Clique problem on a graph of $64$ vertices and graph density $0.5$, we apply Algorithm~\ref{algo:genetic} using the optimized parameters determined in Section~\ref{sec:results_genetic} and $R=100$ generations. Figure~\ref{fig:spatial} shows a section of the qubit connectivity graph (the Chimera graph) of D-Wave 2000Q, where we color the single current best solution of spin reversed qubits with blue (qubit is reversed) and red (qubit is not reversed).

Figure~\ref{fig:spatial} shows that at initialization (left), the assignment is random (with the pre-set probability $p_\text{spin}$ of reversing a bit upon initialization). After $100$ generations, the distribution of spin reversed qubits has changed: As already observed in Figure~\ref{fig:ga_number_reversals}, the proportion of spin reversed qubits has increased.

\section{Conclusions}
\label{sec:discussion}
This paper investigated the spin reversal on the D-Wave 2000Q quantum annealer. In particular, we applied our own approach of a spin reversal which, in contrast to the native one built into D-Wave 2000Q, allows us to specify the probability with which spins are reversed, and we apply the spin reversal to the qubit and chain level. Importantly, we present a genetic algorithm to select the individual qubits which, if spin reversed, yield a modified Ising model that is easier to minimize (resulting in a minimum of better quality).

We summarize our findings as follows:
\begin{enumerate}
    \item The native spin reversal on D-Wave 2000Q seems to be highly optimized for the annealer, even though D-Wave claims that its spin reversal is applied at a constant flipping probability of $0.5$. In our experiments we failed to achieve lower energy solutions in the average case.
    \item We show using a genetic algorithm that selecting the specific spins to be reversed is much more beneficial, and considerably improves upon D-Wave's native spin reversal. Importantly, we show that typically only a handful of generations (around $R=10$) is sufficient to arrive at a combination of individual qubits which, if reversed, yield Ising models with solutions of substantially better quality. Therefore, increasing the computational effort (the number of anneals) by a constant factor (around $R=10$ generations with a population size of $N \leq 80$ each) can yield substantially better solutions.
\end{enumerate}

The genetic algorithm we employ to optimize the spin reversal transform is a rather basic version, and much more sophisticated approaches exist in the literature. Future work includes the implementation and tuning of such advanced methods, with which we hope to (1) find even better sets of qubits to be spin reversed; and (2) decrease the required population size and number of generations, thus resulting in a reduced overhead when optimizing the spin transform.

Moreover, future research is needed to find out about the inner workings of D-Wave's native spin reversal -- being able to reproduce D-Wave's native spin reversal could further improve performance of our genetic algorithm approach.

Additionally, it is always worthwhile to further investigate how the results in this article are dependent on the problem being solved.

\bibliographystyle{plain}

\appendix
\section{Further assessments of Algorithm~\ref{algo:genetic} on chain level}
\label{sec:genetic_qubit}

\begin{figure}
    \centering
    \includegraphics[width=0.49\textwidth]{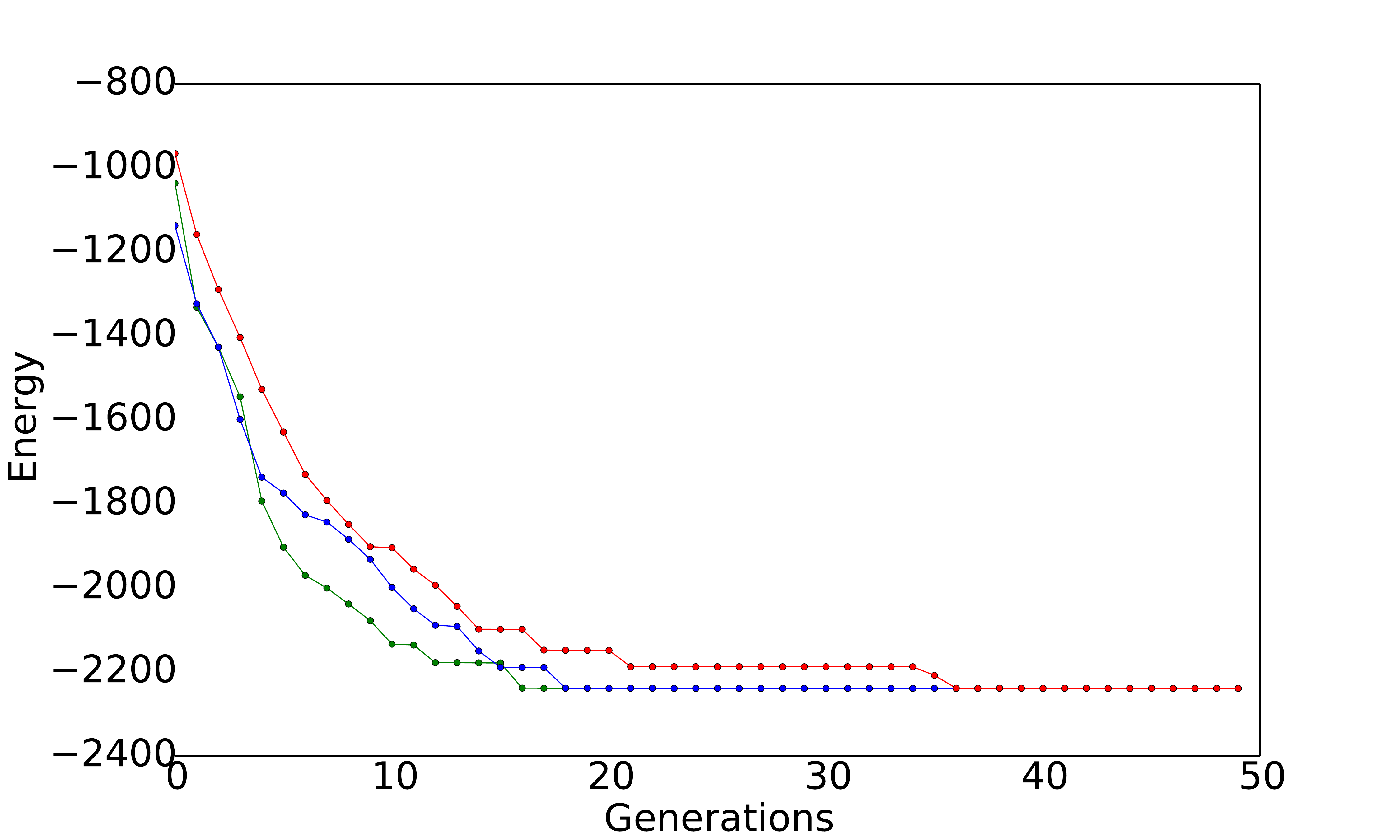}~
    \includegraphics[width=0.49\textwidth]{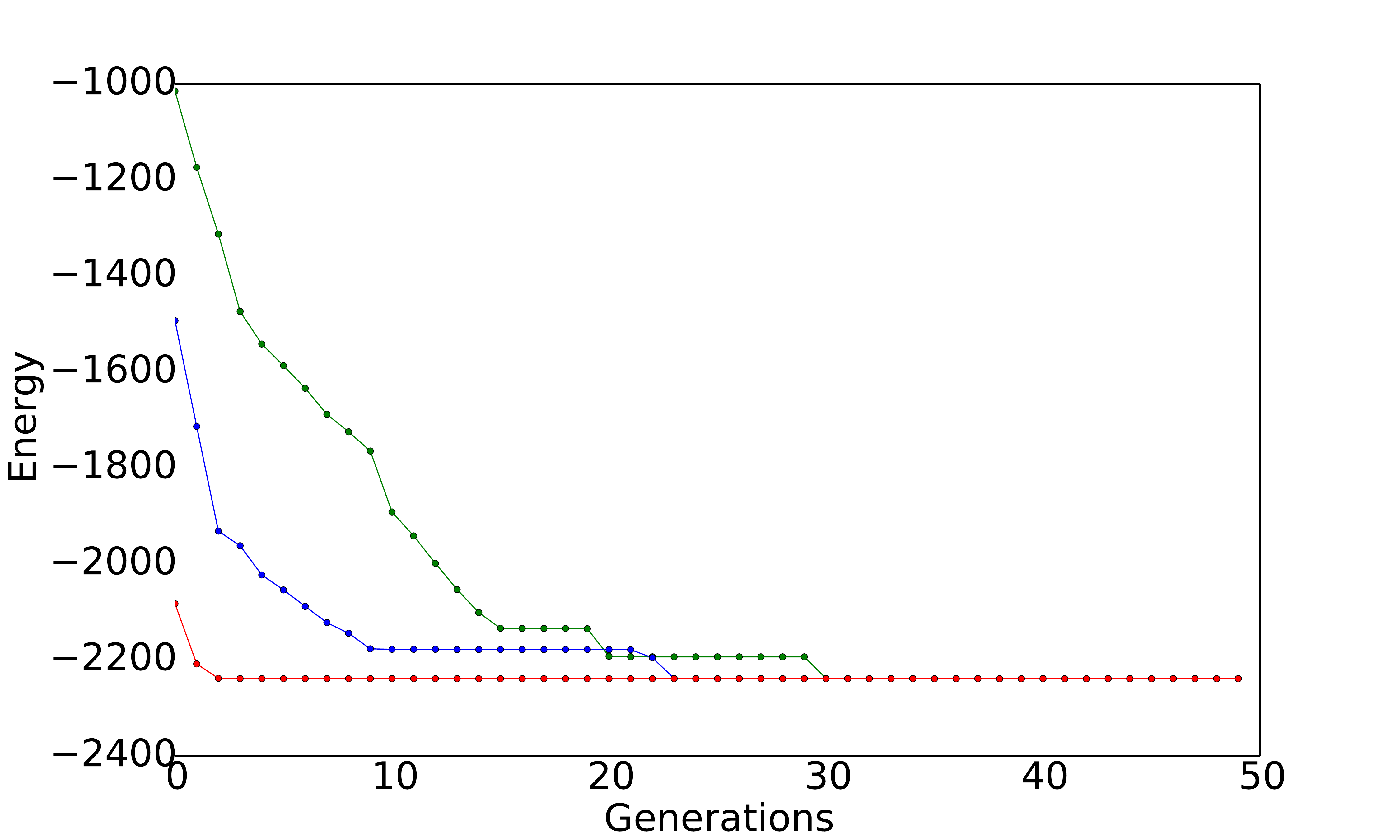}\\
    \includegraphics[width=0.49\textwidth]{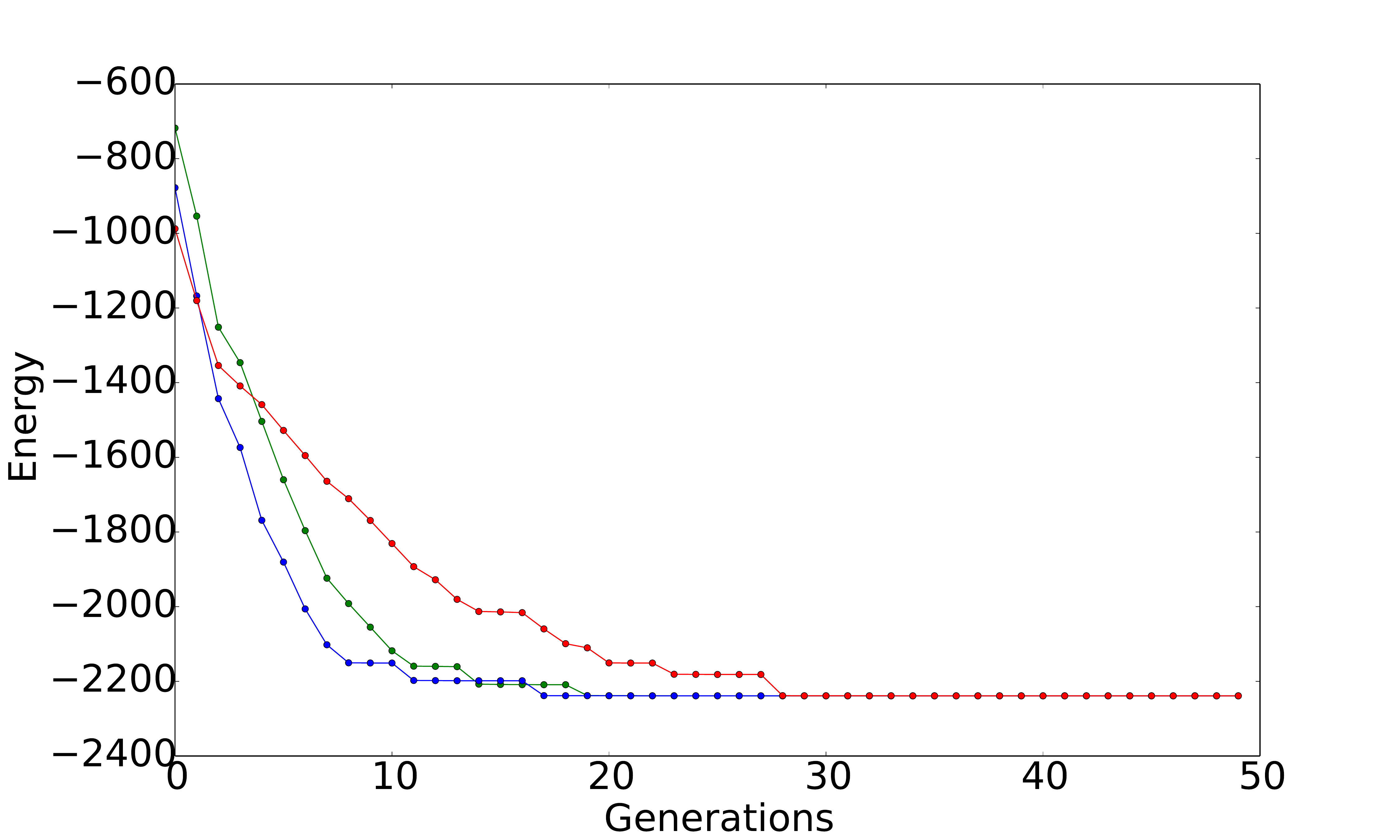}~
    \includegraphics[width=0.49\textwidth]{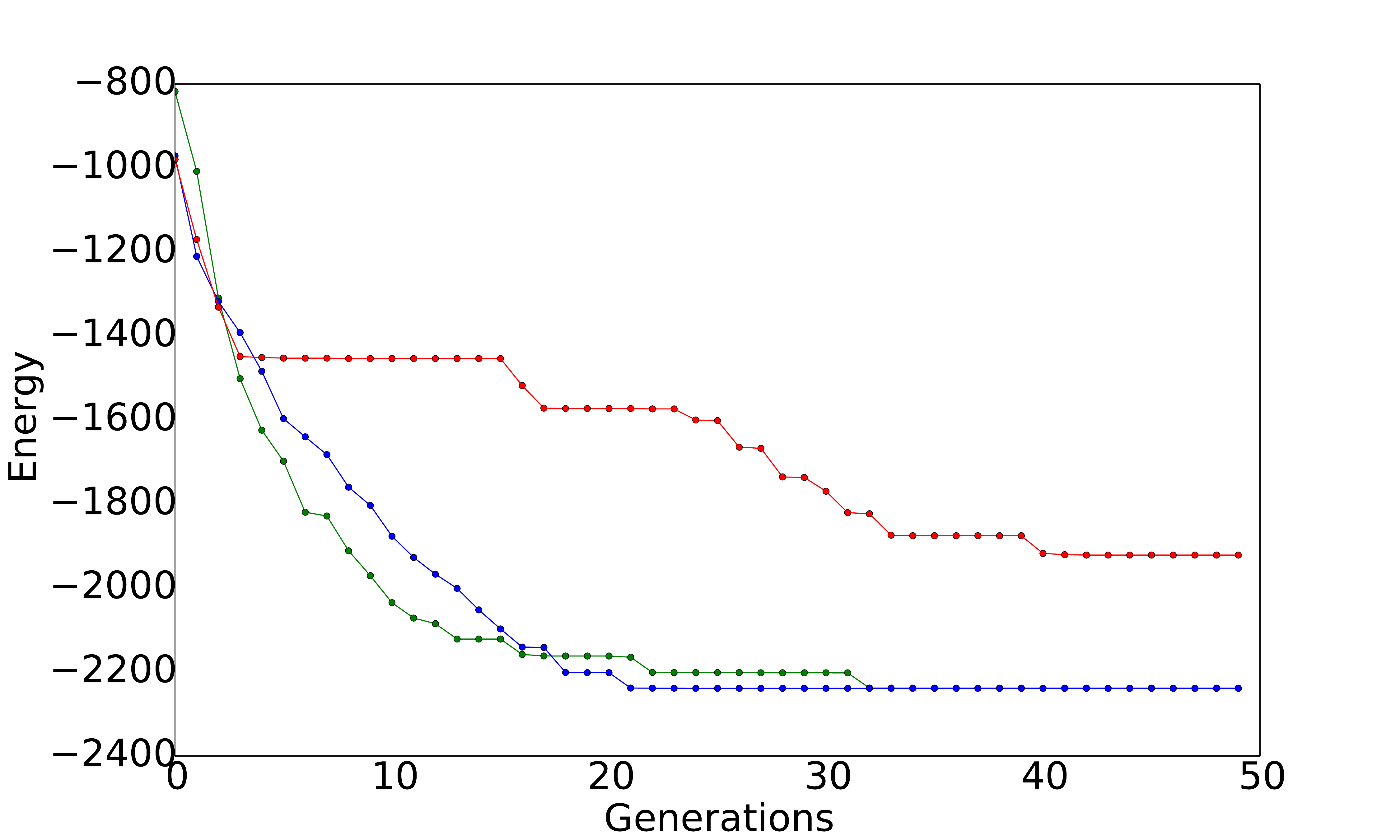}
    \caption{Dependence of Algorithm~\ref{algo:genetic} on its parameters $N \in \{20,50,80\}$ (top left), $p_\text{spin} \in \{0.1,0.3,0.5\}$ (top right), $p_\text{mat} \in \{0.1,0.3,0.5\}$ (bottom left) and $p_\text{mut} \in \{0.001,0.01,0.1\}$ (bottom right) when applied to spin reversal at chain level for 50 generations. We consistently use green for the smallest value, red for the medium value, and blue for the largest value in all subplots.\label{fig:GA_param_qubit}}
\end{figure}

We repeat the assessments of Algorithm~\ref{algo:genetic} in Section~\ref{sec:results_genetic} for the spin reversal applied at chain level using $R$ = $50$ generations. As in Section~\ref{sec:results_genetic}, we evaluate the dependence of the genetic algorithm on its parameters $N$, $p_\text{spin}$, $p_\text{mat}$, and $p_\text{mut}$ while keeping those parameters which are not being varied at their values given in Section~\ref{sec:results_genetic}, except for $p_\text{spin}=0.5$, see section \ref{sec:experiments}. 

We find that Algorithm~\ref{algo:genetic} converges quite quickly to a low energy solution (Figure~\ref{fig:GA_param_qubit}), where the total number of spin reversals is quite low (not shown).

\end{document}